\documentclass[4paper]{article}
\usepackage{jcappubnew}
\usepackage{epsfig,amsmath,latexsym,amssymb}
\newcommand{\be}{\begin{equation}}
\newcommand{\ee}{\end{equation}}
\newcommand{\bea}{\begin{eqnarray}}
\newcommand{\ena}{\end{eqnarray}}
\newcommand{\no}{\noindent}
\newcommand{\nb}{\nonumber}

\renewcommand\o{\omega}

\renewcommand\L{\ensuremath{\Lambda}}
\newcommand\m{\ensuremath{\mu}}
\renewcommand\k{\ensuremath{\kappa}}

\newcommand\n{\ensuremath{\nu}}

\newcommand{\de}{\partial}

\newcommand{\ba}{\begin{eqnarray}}
\newcommand{\ea}{\end{eqnarray}}
\newcommand{\E}{\mathcal{E}}

\title{
FRW Cosmological Perturbations in Massive Bigravity
} 
\author[a,b]{D. Comelli }
\author[c,d,e]{ M. Crisostomi }
\author[d,f]{and L. Pilo }
\affiliation[a]{INFN, Sezione di Ferrara,  I-35131 Ferrara, Italy}
\affiliation[b]{CERN, Theory Division, 1211 Geneva, Switzerland}
\affiliation[c]{School of Physics and Astronomy, University of Nottingham, Nottingham NG7 2RD, UK}
\affiliation[d]{INFN, Laboratori Nazionali del Gran Sasso, I-67010 Assergi, Italy}
\affiliation[e]{Universit\'e Paris Diderot - Paris 7, F-75205 Paris, France}
\affiliation[f]{Dipartimento di Fisica, Universit\`a di L'Aquila,  I-67010 L'Aquila, Italy}
\emailAdd{comelli@fe.infn.it}
\emailAdd{marco.crisostomi@aquila.infn.it}
\emailAdd{luigi.pilo@aquila.infn.it}

\date{\small \today}
\keywords{Massive Gravity, Cosmological Perturbations}

\abstract {
Cosmological perturbations of FRW solutions in ghost free massive
bigravity, including also a second matter sector, are studied in detail.
At early time, we find that sub horizon exponential instabilities are
unavoidable and they lead to a premature departure from the perturbative regime of cosmological perturbations.} 

\begin{document}
\maketitle
\section{Introduction}
Dark Energy is the dominant component of our
Universe, if future observations will establish that its equation of state  differ from the one of a Cosmological Constant
contribution, then 
 we have a case for modifying GR at large
distances and massive gravity can be  a compelling candidate. Great effort was devoted  to extend at the
nonlinear level~\cite{Gabadadze:2010, Hassan:2011vm} the seminal work of 
Fierz and Pauli (FP)~\cite{Fierz:1939ix} and recently a Boulware-Deser (BD) ghost free
theory was found~\cite{Gabadadze:2011, HR}.
Unfortunately, cosmological solutions of the ghost free dRGT theory are
rather problematic: spatially flat homogenous
Friedmann-Robertson-Walker (FRW) solutions simply do not
exist~\cite{DAmico} and even allowing for open FRW solutions~\cite{open}
strong coupling~\cite{tasinato} and ghostlike  
instabilities~\cite{defelice-prl, defelice} develop. In addition the
cutoff of the theory is rather low~\cite{AGS}, namely $\Lambda_3=\left(m^2 \,
  M_{Pl} \right)^{1/3}$. For a recent review see \cite{de-rham-long,defelice-rev}.

A possible way out is to give up Lorentz invariance and requires only rotational
invariance~\cite{Rubakov,dub,usweak}. Within the general class of
theories which propagate five DoF found in~\cite{uscan,uslong}, in
the Lorentz breaking case most of the theories have much safer
cutoff $\Lambda_2 =(m \, M_{Pl})^{1/2}\gg \Lambda_3$ and also avoid all
of the phenomenological difficulties mentioned above, including the
SWtroubles with cosmology~\cite{cosmogen}.
Another option is to promote the nondynamical metric entering in the
construction of massive gravity theory to a dynamical
one~\cite{DAM1,PRLus} entering in the realm of bigravity originally introduced by Isham, Salam
and Strathdee~\cite{Isham}.

In the bigravity formulation FRW homogenous solutions
do exist~\cite{uscosm,hasscosm,russ}, however cosmological perturbations, for modes inside the horizon,
start to grow too early and too fast when compared with GR, as a 
result the linear regime becomes problematic already during the
radiation/matter era~\cite{uspert}.  
The reason of such peculiar behaviour of the scalar perturbations could
be {\it naively} traced back  to the  FRW background solution 
which is controlled by the parameter $\xi$  (the ratio of the
conformal factors of the two metrics)   and to the absence of matter
coupled to the second metric whose pressure could support inside horizon gravitational perturbations. 

In presence of only ordinary matter, coupled with the first metric,
{\it only} small values of the parameter  $\xi$ give an acceptable early
time cosmology. The introduction of the second matter component
provides other consistent background solutions where the values of
$\xi$ can be also of order~1 and, at the same time, provides the necessary pressure support to infall perturbations.

 So in this paper we will extend our  previous analysis to the case where an additional  matter sector
is minimally coupled to the second metric.
Though we do not consider the problem, the second matter sector
could be also relevant for dark matter~\cite{Yuk1,Yuk2}.

The outline of the paper is the following: in section \ref{bi} we
review the bigravity formulation of massive gravity and the extension
to the case where a second matter sector is present; in section
\ref{frwsec} we study FRW solutions and cosmological perturbations are
analysed in section \ref{pert-sect}.

\section{Massive Gravity and Bigravity}
\label{bi}
Any modification of GR that turns a massless graviton into a massive
one calls for additional DoF. 
Basically, GR is
deformed by a scalar function $V$ built from the physical metric $g$ that
couples with matter and a second metric $\tilde g$.
Besides phenomenological issue, dealing
with a non dynamical metric is rather awkward, in this context it is
natural to promote the second metric to a fully dynamical field, see for instance~\cite{myproc}. 
Thanks to  $\tilde g$, it is possible to build nontrivial 
diffeomorphism invariant terms without derivatives of the metric.
Expanding the metric around a fiducial background, such terms  
lead precisely to the mass term for the graviton.
Consider the action
\be
S=
\int d^4 x \left\{ \sqrt{\tilde g} \, \left[  \kappa   \; M_{pl}^2\; \tilde R +   L_{\tilde{\text{M}}} \right] +\sqrt{g} \;\left[ 
M_{pl}^2 \;\left( R
-2 \, m^2   \, V \right)  + L_{\text{M}} \right] \right\},
\label{act} 
\ee
where $R$ and $\tilde R$ are the corresponding Ricci scalars and the deforming 
potential $V$ is a scalar function of the tensor $X^\mu_\nu = {g}^{\mu
\alpha} {\tilde g}_{\alpha \nu}$. Ordinary matter is minimally coupled
with $g$ and is described by $L_{\text{M}}$. In order try to cope with the
instabilities found in~\cite{uspert} we shall introduce a second
matter sector that couples minimally with $\tilde g$ and it is
described by  $ L_{\tilde{\text{M}}} $. 
The constant $\kappa$ controls
the relative size of the strength of gravitational interactions in the
two sectors, while $m$ sets the scale of the graviton mass. In
particular, in  the limit $\kappa \to \infty$, the second metric gets
frozen to a prescribed background value. Removing the second matter
sector one recovers the previously studied bigravity theories, see for instance~\cite{ussphe}.

The modified Einstein equations can be written as\footnote{When not
specified, indices of tensors related with $g$($\tilde g$) are raised/lowered with
$g(\tilde g)$}
\begin{gather}
\label{eqm1}
\,{E}^\mu_\nu  +  Q_1{}^\m_\n = \frac{1}{2\;M_{pl}^2 }\, {T}^\mu_\nu \,,  \\
\label{eqm2}
\kappa  \, {\tilde E}^\mu_\nu  +  Q_2{}^\m_\n = \frac{1}{2\;M_{pl}^2 }\, {\tilde T}^\mu_\nu \;;
\end{gather}
where we have defined $Q_1$ and $Q_2$ as effective energy-momentum tensors induced by the
interaction term.
The ghost free
potential~\cite{Gabadadze:2011, Hassan:2011vm}\footnote{A very similar potential
  having the same form but with $X$ instead of $X^{1/2}$ was
  considered in~\cite{usex}.}  $V$ is  a special  scalar function 
of $Y^\mu_\nu=(\sqrt{X})^\mu_\nu$ given by
\be
\begin{split}
& V=\sum_{n=0}^4 \, a_n\, V_n \,,\qquad n=0\ldots4 \, , \qquad \tau_n
= \text{Tr}(Y^n); \\
&V_0=1\,\qquad 
V_1=\tau_1\,,\qquad
V_2=\tau_1^2-\tau_2\,,\qquad
V_3=\tau_1^3-3\,\tau_1\,\tau_2+2\,\tau_3\,,\\[1ex]
&V_4=\tau_1^4-6\,\tau_1^2\,\tau_2+8\,\tau_1\,\tau_3+3\,\tau_2^2-6\,
\tau_4\, .
\end{split}
\label{eq:genpot}
\ee 
In \cite{HRBI} it was shown that in the bimetric
formulation the potential $V$ is BD ghost free. We have that 
\bea
\label{eq:q1}
 Q_1{}_\nu^\mu &=&  { m^2}\, \left[ \;  V\; \delta^\mu_\nu \,  - \,  (V'\;Y)^\mu_\nu  \right]\\[1ex]
\label{eq:q2}
 Q_2{}_\nu^\mu &=&  m^2\, q^{-1/2} \, \; (V'\;Y)^\mu_\nu ,
\ena
where  $(V^\prime)^\mu_\nu = \de V / \de Y_\mu^\nu$ and  $q =\det
X=\det(\tilde g)/\det(g)$.  

The canonical analysis~\cite{HRBI, GF} shows that in general 7 DoF propagate; around a
Minkowski background, 5 can be associated to a massive spin two
graviton and the remaining 2 to a massless spin two graviton.  
We consider only the case where each matter sector is minimally
coupled with  only its own metric field. 
Allowing the second metric to couple also with
standard matter would result in a violation of the equivalence
principle; indeed, it is not possible to locally put both
metrics in a Minkowski form.

\section{FRW Solutions in Massive Bigravity}
\label{frwsec}
Let us consider FRW background solutions in massive bigravity of the form 
\be
\begin{split}
ds^2 &=   a^2(\tau) \left(- d \tau^2 +   dr^2 + r^2 \, d \Omega^2 \right) =
 {\bar g}_{1 \, \mu \nu} dx^\mu dx^\nu \,, \\
\tilde{ds}^2 &= \omega^2(\tau) \left[- c^2(\tau) \, d \tau^2 + dr^2+  r^2 \, d
  \Omega^2 \right] =
 {\bar g}_{2 \, \mu \nu} dx^\mu dx^\nu\, .
\label{frw}
\end{split}
\ee
It is convenient to define the standard Hubble 
parameters for the two metrics and the ratio between the two scale factors
\be
{\cal H} =\frac{d a}{d\tau}  \frac{1}{a} \equiv \frac{a'}{a}%= H \, a  
\, ,
\qquad {\cal H}_\o \equiv \frac{\o'}{\o}=\frac{\xi'}{\xi}+{\cal H} %= H_\o \, \o
 \, , \qquad \xi \equiv\frac{\o}{a} \, ;
\label{hubble}
\ee
where with $'$ we always denote the derivation with respect to the conformal
time $\tau$.
Solutions fall in two branches depending on how the covariant
conservation of $Q_{1/2}$, enforced by the Bianchi
identities, is realized. It turns out that the physically interesting
case~\cite{uscosm,uspert} is when, as a consequence of the
conservation of $Q_{1/2}$, we have that 
\be
c = %\frac{H_\o}{H} \, \xi = 
 \frac{{\cal H}_\o}{{\cal H}} \,,\;\;\;\;\;\; \xi'=(c-1)\;{\cal H}\;\xi 
\;\;\;\;  {\rm with}\;\;\; c>0 \, .
\label{bnew}
\ee 
We will not discuss the other branch of solutions where $\xi$ is constant
and the effect of gravity modification amounts to a cosmological
constant and  perturbations are strongly coupled~\cite{uspert}, as  expected.

The expansion rate follows from  the equation
\be
\frac{3 \,\mathcal{H}^2}{a^2}= 8 \pi  G \, \rho_1  + m^2\; \left(6\, a_3 \,
  \xi ^3+6 \,a_2 \,  \xi^2+3\, a_1 \,  \xi +a_0 \right) \, .
\label{hh}
\ee
The presence of the second metric is equivalent, for the first sector, to a gravitational
fluid with energy density $\rho_g$ given by
\be
\rho_g =\frac{m^2 \left[  6 \, \xi^2  \left(a_3 \, \xi
   +a_2\right)+ 3 \, a_1 \,  \xi+a_0\right]}{8 \pi  G} \, \, ;
\ee
with an equation of state $p_g=w_g \, \rho_g$ of the form
\be
w_g =-1- \frac{\left(6 \, a_3\, \xi ^2+4 \, a_2 \, \xi +a_1\right)
  \xi'}{\mathcal{H} \;\left[ 6 \, \xi^2 \, \left(a_3\, \xi
   +a_2\right)+ 3 \, a_1 \,  \xi+a_0\right]} \, .
\ee
The conservation of energy-momentum tensor for the two fluids leads to 
\bea
\label{mat12}
\rho_1'+3\;{\cal H}\;(\rho_1+p_1)=0,\qquad \rho_2'+3\;{\cal H}_\o\;(\rho_2+p_2)=0,
\ena
thus \;for $p_i  = w_i \,
 \rho_i$  we \;have $
\rho_1  =
\rho_{1}^{\text{in}} \, a^{-3(w_1+1)}$ and $ \rho_2  = \rho_{2}^{\text{in}} \,
\o^{-3(w_2+1)}$.  

Finally, using (\ref{bnew}) in the time-time component of the Einstein equations for
the second metric we get that the ratio $\xi$ of the two scale parameters satisfies the following algebraic equation
\be
\xi ^2 \left(\frac{8\; a_4}{\kappa }-2\; a_2\right)+\xi 
   \left(\frac{6 \; a_3}{\kappa }-a_1\right)+\frac{a_1}{3 \; \kappa \, 
   \xi }+\frac{2 \; a_2}{\kappa }-2\;  a_3 \, \xi ^3-\frac{a_0}{3}
=\frac{8 \pi G}{3 \, m^2}\left( \rho_1 - \frac{\xi^2 \rho_2}{\kappa} \right)  \, .
\label{conII}
\ee
The analysis  is identical when the {\it same}  spatial
curvature $k_c$ is introduced in (\ref{frw}) for both metrics\footnote{The spatial
curvatures must be equal for consistency~\cite{uscosm}.}.
The presence of the second matter opens the possibility for a 
behaviour of $\xi$ different from the one found in \cite{uscosm}.
 
We  assume that the mass scale $m$ is related to the present
 cosmological constant as $\; m^2\,M_{pl}^2\propto \Lambda$ and the
 equation of state for matter one and two is  such that $w_{1,\,2}>
 -1$. The assumption   on the scale   $m$ is natural if massive gravity
 is relevant for  the present acceleration of the Universe
 \footnote{We do not consider here the case  \cite{DeFelix} $m^2\;M^2_{pl}
 \gg\rho_1$, where the scale of $m$ is not related with the present
 acceleration of the Universe.}.
 In order to not spoil early cosmology (say before nucleosynthesis till after the decoupling time), the contribution 
proportional to $m$ in (\ref{hh}) have to kick in only at small
redshift ($z\sim 10$) when ``dark energy'' starts to dominate the expansion rate. This
is the case when  
\be
\frac{3 \mathcal{H}^2}{a^2}\simeq  8 \pi  G \, \rho_1\quad {\rm
  implying }\quad m^2\; \sum_{i=0}^3\;(a_i\;\xi^i)  \ll  8 \pi  G \,
\rho_1 \, ,
\label{early}
 \ee
or equivalently
\be
\label{first}
\frac{\Lambda}{\rho_1} \, \sum_{i=0}^3 a_i \, \xi^i \ll1 \, .
\ee
Now, for most of the history of our Universe (matter and radiation periods) $\rho_1 \gg \Lambda$, thus
(\ref{first}) is naturally satisfied unless $\xi$ evolves to values of $\sim \rho_1/\Lambda$.  
As a result, in such a regime,  the implementation of eq.(\ref{first})
in eq.(\ref{conII}) requires that at the leading order
\be \label{chi}
8\,\pi\,G\,(\rho_1\,\kappa-\rho_2\,\xi^2) \simeq
\begin{cases}
{\Large \frac{a_1\, m^2}{\xi }} & \qquad \text{when } \xi \ll 1\\[.3cm]
\quad 0 & \qquad \text{when } \xi \sim 1
\end{cases} \,\,\, .
 \ee
In absence of a second matter sector, the solution $\xi \sim 1$ could not exist.
Of course, when (\ref{early}) holds, while the dynamics of $a$ is not affected by $\xi$, on the contrary, the impact on ${\cal H}_\omega$
can be relevant, see (\ref{hubble}).
According to eq.(\ref{chi}), the following regimes for the  background
value of $\xi$ emerge  
\begin{itemize}

\item[({\bf A})] When $\xi^2\;\rho_2\gg \rho_1\gg \L$ \footnote{Notice that,
    being $\xi^2\gg \frac{\L}{\rho_2}\simeq \xi^3$, then $\xi\ll1$ and so we are in the region where $\rho_2\gg\rho_1\gg\L$.}
\be
\xi\simeq  -\left( \frac{a_1\,m^2 }{8\,\pi\,G\;\rho_2
  }\right)^{1/3}\propto \frac{\L^{1/3}}{\rho_2^{1/3}}\ll\frac{\L}{\rho_1}\ll 1,\qquad
\qquad \text{with } 
c \simeq -\frac{1}{w_2}  \, .
\ee
The above expression can be rewritten also in the form 
\be
\xi=\left(-\frac{8\,\pi\,G\,\rho_2^{\text{in}}}{a_1\;m^2}\right)^{\frac{1}{3\,w_2}}\;a^{-\frac{1+w_2}{w_2}}
\, ,
\ee
where the explicit  time dependence of  $\xi$ is shown. The above
expressions are valid when $w_2 <0$. Clearly, we have that  $c >0$ and
we also need $a_1 <0$ so that $\xi$ is
real and positive. Being $\xi \ll1$, (\ref{early}) is
satisfied. Requiring $w_2 <0$ is rather exotic, nevertheless, as will be
shown in section \ref{caseA}, it does not help to avoid instabilities.

\item[({\bf B})]  When  $\rho_1\gg\xi^2\;\rho_2$   and at any time
  $\rho_1\gg \L$. 
  
 This case was considered in~\cite{uspert} when a
  single matter sector was present. Clearly (\ref{early}) is easily
  satisfied. The value for $\xi$ is of the form 
\be
\xi\simeq \;\frac{a_1\,m^2}{8\,\pi\,G\;\rho_1\;\kappa}\propto
\frac{\L}{\rho_1}\ll1, \qquad \qquad 
\text{with } \; c\simeq (4+3\,w_1) \, .
\ee
and self consistency requires that 
\be
\rho_2\ll \frac{\rho_1^3}{\L^2} \, .
\ee

\item[({\bf C})] When $ \rho_1 \simeq \xi^2 \; \rho_2 \gg
  \Lambda$
\be
\xi\simeq \left(\kappa\; \frac{\rho_1}{\rho_2}  \right)^{1/2}
%\Rightarrow \xi \simeq
= \,
\xi_{in}\;a^{\frac{3\,(w_1-w_2)}{1+3\,w_2}}
,\qquad \qquad \text{with }
c\simeq \frac{1+3 \;w_1}{1+3\; w_2} \, ;
\ee
where we used the solutions of eq.(\ref{mat12}) and   $%r_\rho 
\xi_{in}= (\k\,\rho_1^{\text{in}}/\rho_2^{\text{in}})^{-1/(1+3\,w_2)}$ defines the initial time  conditions  in terms of the initial density ratio.
In such a regime  $\rho_2\propto
a^{-3\frac{(1+w_2)\,(1+3\,w_1)}{1+3\,w_2}}$, thus only when
$w_2>-\frac{1}{3}$ matter density in the second sector decreases with time, while
$\xi$ can grow or decay depending on the sign of $(w_1-w_2)$.
When $w_2 > w_1$, going back in time, $\xi$
grows; nevertheless, condition (\ref{first}) is still satisfied if $w_1\ge0$.
The validity region of such an approximated solution is in the range
\be\label{cc}
\frac{\L}{\rho_1}\ll \xi\ll \left(\frac{\rho_1}{\L}\right)^{n} \; ,
\ee
where the power $n$ can be $1/3,\,1/2 $ or $1$ depending on the $a_i$ values, see \cite{uscosm} for details.
%, see footnote \ref{foot}.
When $w_1 > w_2$ and $\xi$
decreases going back in time, the above lower bound holds  for  $w_2 > -1/(4+ 3 w_1)$.

\item[({\bf D})] When $\rho_2=0$, also the
case of very large $\xi$ is possible, with
\be
\xi \propto \left(\frac{ \rho_1}{\L}\right)^n \, ,
\ee
which gives $c<0$. The power $n$ is the same as in eq.(\ref{cc}), see \cite{uscosm}.
 Thus,  not only (\ref{early}) is violated but also $c$ is
negative. Starting from a negative $c$ in order to get to  a quasi dS
phase, where $c \sim 1$, one has to cross $c=0$ where $\tilde g$  is
singular.\footnote{This point was overlooked in~\cite{hasscosm, Berg, kov}. We only
consider FRW-like backgrounds where $c>0$.} 
\end{itemize}

\medskip

\no Finally, looking at the validity of our
approximation, we found that the explored range of the $\xi$ values can be divided in the following 
 disjoined  regions
 \be
 \xi_{(\bf A)}\ll\; \xi_{(\bf B)}\sim\frac{\L}{\rho_1} \;
 \ll\xi_{(\bf C)}\ll \;
 \xi_{(\bf D)}\sim \left(\frac{ \rho_1}{\L}\right)^n \, ,
 \ee
which cover the whole range of $\xi $; except $(\bf{D})$, all  cases
are compatible with eq.(\ref{early}), i.e. an early time standard 
FRW universe.

\section{Perturbed FRW Universe}
\label{pert-sect}
Perturbations around the solution (\ref{frw}) can be studied along the
same lines of~\cite{uspert}. We focus here on the scalar sector; in
the vector and tensor ones, the results are very similar to the case
with only $\rho_1$ and they can be found in~\cite{uspert}.  
In the scalar sector we have 8 fields and two independent gauge
transformations, as a result we can form 6 independent gauge invariant
scalar combinations $ \Psi_1, \, \Psi_2, \, \Phi_1, \, \Phi_2, \,
{\cal E}, \, {\cal B}_1$ for the metric perturbations. For matter
we have the gauge invariant density perturbations $\delta
\rho_{1/2 \, \text{gi}}$ and the scalar part of velocity perturbations $\delta
u_{s \, 1/2}$. The various definition can be found in Appendix \ref{pert-app} where also the full set of equations is given. 

The fields  ${\cal B}_1$
and $\Psi_{1/2}$ are non dynamical and can be expressed in
terms of ${\cal E}$ and $\Phi_{1/2}$, in particular
\be
\Psi_1 +\Phi _1=  m^2  \, a^2 \, f_1 \, \mathcal{E} \, , \qquad \qquad
 \Psi_2
+\Phi _2 =-\frac{m^2  \, a^2 \, f_1 \, \mathcal{E} 
}{
   \kappa \;c\; \xi^2 }\,  ;
\label{psi}
\ee
where $f_{1/2}$ are defined in eq.(\ref{fdef}).
The fields  ${\cal E}$ and $\Phi_{1/2}$ satisfy three second order equations; thus 3 scalar DoF propagate. 

The condition (\ref{early}) guarantees only that background solution
follows closely GR cosmology with standard matter (sector 1) until the present
epoch. Of course we need more than that: we need to be sure that
perturbations, in particular the ones related to the new degrees of
freedom,  do not start growing too early. Indeed, that is precisely what
happen when only ordinary matter is present: at early time, the mode  $\Phi_2$
inside the horizon grows exponentially, though $\Phi_1$ and $\delta=
\delta \rho_{1 \, \text{gi}}/\rho_1$ are the same as in GR. As result we have
to face a very early breakdown of perturbation theory. Apparently,
this point was not taken into account  fitting the
parameters $a_i$ and $m$ against observations~\cite{amendola}. 
Basically, in the presence of the
aforementioned instabilities, structure formation will be completely
different. Thus, a preliminary necessary condition is to get rid of
exponential instabilities, irrespective of their tachyonic or ghost
nature. 
In what follows we will show that also the presence of a second matter
sector is not instrumental to avoid such a kind of instabilities. 

\subsection{Structure of the evolution equations}
The equations are rather complicated, however at early
times we can expand using the small parameter  $\epsilon = m \;
{\cal H}^{-1}\sim (\Lambda/\rho_1)^{1/2}$. 
 Formally this is equivalent to expanding the equations of the
perturbation for small $m$. We stress that in the $m \to 0$ limit
there is no guarantee to recover GR as discussed in details in~\cite{uspert}.

In all   cases ({\bf A}), ({\bf B}) and ({\bf C}), $\Phi_1$, at leading order in 
$\epsilon$, satisfies the following equation 
\be
\Phi _1'' +\frac{6\left(w_1+1\right)}{(1+3 \, w_1)\;\tau}\Phi _1'+k^2\; w_1\;
\Phi _1=0 \, ;
\label{eq1}
\ee
that coincides with the one in GR. In the radiation epoch,
sub-horizon modes oscillates, dumped by a factor $a^2$, while super 
horizon modes are frozen and $\Phi_1=$ constant. In a matter
dominates Universe  $\Phi_1$ is always  constant.  Thus, at leading
order in $\epsilon$, the dynamics of $\Phi_2$ and $\E$ is  described
by a system of coupled second order ODEs of the form
\be
\phi'' + {\cal D} \, \phi' + {\cal M} \, \phi + z_1 \, \Phi_1 + z_2 \, 
\Phi_1' \,  =0  \, , \qquad \phi = \begin{pmatrix} \Phi_2 & \\ \E_N \equiv \E/\tau^2  \end{pmatrix} \, .
\label{dyn}
\ee
where ${\cal D}$ and ${\cal M}$ are suitable 2$\times$2 matrices and
$z_{1/2}$ functions of $\tau$ and $k$. We have also conveniently introduced a
dimensionless field $\E_N= {\cal E}/\tau^2$. Thus, once $\Phi_1$ is found from
(\ref{eq1}), it enters in (\ref{dyn}) as a source term. As shown in
appendix \ref{details}, the eqs (\ref{dyn}) correspond, for sub horizon and super horizon modes, to a coupled system of
Bessel-like equations.
It turns out that for cases ({\bf A}) and ({\bf C}), the system (\ref{dyn})  further simplifies
because  the dynamics of $\Phi_2$ decouples from the one of $\E_N$ and
stability can be established simply studying the mass term. For the case
({\bf B}),  on the contrary, one has to do a more involved  analysis.

\subsection{Case ({\bf A})}
\label{caseA}
One has to be careful in the expansion; indeed here one
can expand for small $m$ only if $w_2< -1/3$.  The result, this
time, is that  also the equation for $\Phi_2$ is decoupled.  In particular, we have that
\be
\Phi _2'' +
\frac{6 \left(1-|w_2|\right)
   }{|w_2|\, \left(3\, w_1+1\right)\, \tau} \, \Phi _2'    -
   \left(\frac{k^2}{|w_2| }+\frac{|w_2| \,\left( 3 \,w_1+4\right)-4}{ w_2^2 \,\tau^2\,\left(3 \,w_1+1\right)^2}
   \right) \;\Phi _2 =0 \, .
\ee
The above equation can be easily solved in terms of Bessel
functions. However, it is clear that the solution has an exponentially
growing mode. Indeed, inside the horizon $x=k \,\tau \gg 1$,  the mass term is simply proportional to $-|w_2|$ and is
negative. The solution reads
\be
\Phi_2 = \left(x |w_2|\right)^{2-\frac{3}{2 |w_2|}} \left[\alpha_1 \,
    J_\nu \left( \frac{-i \, x}{\sqrt{|w_2|}}\right) + \alpha_2 \,
    Y_\nu\left( \frac{-i \, x}{\sqrt{|w_2}|} \right)
  \right] \, , \qquad \nu =\frac{\sqrt{4 w_2 \left(4
        w_2+1\right)+5}}{2 |w_2|} \, ,
\ee
Clearly $\Phi_2$ grows like $e^{x/\sqrt{|w_2|}}$.
The same instability is present also for the field $\E$ whose mass term, inside the horizon, gets the value
$\frac{k^2\left( w_1 -1 \right)}{|w_2| \left(3w_1+1\right)}$\,.
As a result, exponential instabilities are always present in both $\Phi_2$ and $\E$ and the background ({\bf A})
is pathological. The behaviour of super horizon modes is similar to
the case ({\bf B}), discussed bellow.

\subsection{Case ({\bf B})}
%\label{caseA}
As for the case ({\bf A}),  the  expansion for small $m$ is a bit tricky, indeed $\xi\simeq \;\frac{a_1\, m^2}{8\,\pi\,G\;\rho_1\;\kappa} $
goes to zero when $m \to 0$ and all quantities must be expanded to next
to leading order. 
In this case, as shown in appendix~\ref{details}, the equations for
$\Phi_2$ and $\E$ stay coupled. The only way to decouple them is to
work with a forth order equation for one of the two field. Taking for
simplicity $w_1=1/3$, we get for \underline{\it sub horizon modes}
\be
\begin{split}
&{\cal E}_N{}^{(4)} +\frac{5
 \left(3 w_2+5\right)}{\tau} \,  {\cal E}_N^{(3)}
+k^2\;\left(25\;w_2-\frac{5}{3}\right)\; {\cal E}_N'' 
  +k^2\;\frac{25\left(  9\;w_2-1\right)  }{\tau} \, {\cal E}_N'
  -k^4\;\frac{125 \, w_2}{3}\;{\cal E}_N=0 \, .
\end{split}
\label{high1}
\ee
and for $w_1=0$
\bea
\mathcal{E}_N^{(4)} &+&\frac{8\, \left(3\,
   w_2+5\right) \mathcal{E}_N^{(3)}}{\tau}+k^2 \left(16\, w_2-1\right) \mathcal{E}_N^{(2)}
+\frac{8\, k^2 \,\left(29\, w_2-3\right)
  }{\tau}\, \mathcal{E}_N'-k^4\,16 \, w_2 \,\mathcal{E} + \frac{k^2
    (20\, w_2+3)}{\tau} \Phi_1' \nb\\
&+&  \frac{8 \, k^4 \, w2}{3} \, \Phi_1=0 \, .
  \label{high2}
\ena
Even before attempting solving  (\ref{high1}) and (\ref{high2}) one
see that an exponential instability is expected. Indeed, for ${\cal D}$
and ${\cal M}$ in (\ref{dyn}) we have that
\be
\begin{split}
&\text{Det}({\cal D})=\frac{24 \left(3 w_1+4\right){}^2 \left(w_2+1\right)}{\tau^2
\,    \left(3 w_1+1\right)^2} \, , \qquad \quad \; \; \; \; \; \text{Tr}({\cal D})=\frac{2 (4 + 3
 w_1) (5 + 3 w_2)}{\tau(1 + 3  w_1)} \, ,  \\ 
&\text{Det}({\cal M})= -k^4 \left(2 w_1+1\right) \left(3
  w_1+4\right)^2 w_2\, , \qquad \text{Tr}({\cal M})=k^2 \left[ (4 + 3
  w_1)^2 w_2  - 2 w_1-1 \right] \,.
\end{split}
\ee
Thus, while ${\cal D}$ is positive definite, ${\cal M}$ has at least a
negative eigenvalue; in particular the eigenvalues of ${\cal M}$ are
given by
\be
\lambda_1 = - k^2 \, (2 w_1 +1) \, , \qquad \qquad \lambda_2 =k^2
\left(3 w_1+4\right)^2 w_2 \, .
\ee
Clearly, the fact that $\lambda_1 <0$ will lead to  an 
exponential growth of sub horizon modes. It should be stressed that $\lambda_1$ does not depend
on $w_2$ and precisely coincides with the negative mass term of $\E_N$
found in the case where a single matter sector was
present~\cite{uspert}\footnote{For reference, when  $\rho_2 = 0$, $\Phi_2$
  has a tachyonic mass equal to $\lambda_1$ and $\E_N= -\frac{2}{3}
  \Phi_2$, for  $w_1=1/3$.}. 
The numerical solution of (\ref{high1}) and (\ref{high2}) confirms
that  there is no value of $w_2$ such that ${\cal E}_N$ does not grow
exponentially. It is evident that sub horizon instability cannot be
avoided.  

For \underline{\it super horizon modes} we can gives directly the full solutions%,neglecting decaying modes and taking $w_1=1/3$ 
\bea
\mathcal{E}= \bar{\mathcal{E}}_1 \,\tau^{-15 w_2-1}+\bar{\mathcal{E}}_2\,\tau^{-\frac{9}{2}-\frac{\sqrt{21}}{2}}
   +\bar{\mathcal{E}}_3\,\tau^{\frac{1}{2} \left(\sqrt{21}-9\right)}
   +\frac{\bar{\mathcal{E}}_4}{\tau^7}-\frac{32
      \,\tau^2\, \bar \Phi _1}{37}\quad{\rm for}
\quad w_1=\frac{1}{3}\\
 \mathcal{E}=   \bar{\mathcal{E}}_1\, \tau^{-24 \,w_2-4}+\bar{\mathcal{E}}_2\,\tau^{-\frac{15}{2}-\frac{\sqrt{33}}{2}}
   +\bar{ \mathcal{E}}_3\,\tau^{\frac{1}{2} \,\left(\sqrt{33}-15\right)}
   +\frac{\bar{\mathcal{E}}_4}{\tau^{13}}-\frac{21 \,\tau^2\, \bar \Phi_1}{82}\quad{\rm for}\quad w_1=0
\ena
%
%(radiation) for simplicity, 
where the $ \bar{\mathcal{E}}_i$ are the values of $\E$ at same initial time and $\bar
\Phi_{1}$ is the  frozen value of $\Phi_1$. Notice that in
particular for $w_1=\frac{1}{3}$ the non decaying mode of the three perturbations are
\be
\Phi_1= \bar \Phi_1 \, \qquad  \Phi_2= \frac{39}{37} \, \bar \Phi_1 \,
, \qquad \E  = - \frac{32}{37} \,\tau^2 \,  \bar \Phi_1 \; .
\ee
In the metric perturbations actually 
$\E$ enters  with the combination $k^2
 \, \E \propto (k\;\tau)^2 \ll1$, as a result it  stays very small and there are no
 consequences on the validity of the perturbative expansion.
In addition, from (\ref{psi}) we get that
\be
\Psi_1 + \Phi_1 \approx 0 \, , \qquad \qquad \Psi_2 + \Phi_2 \approx
\frac{96}{185} \, \bar \Phi_1 \, .
\ee
Thus, perturbations in the sector one, relevant for our matter,
are indistinguishable from GR at early times. In the second sector the two
Bardeen potentials are not equal even if the source is a perfect fluid.

\subsection{Case ({\bf C})} 
As shown in Appendix \ref{details}, also in this case the dynamics of $\Phi_2$ is decoupled and its
equations are similar to the ones in case ({\bf A}).
Inside the horizon, simply looking at the time dependent mass terms we find that they are positive, avoiding
instabilities, when (see appendix \ref{details})
\be
w_{1/2} >0  \, , \qquad \qquad 3\, w_1+1-\frac{4 \,f_1}{f_2}>0 \, ,
\label{condC}
\ee
where $f_{1,\,2}$, see eq.(\ref{fdef}),  are $\tau$ dependent. Notice that when $f_1=f_2$ the above condition cannot be
satisfied if $0 \leq w_1 <1$.  Actually,  we have  $f_1=f_2$  when $c=1$ and/or $a_2=a_3=0$ (as in the simplest bigravity model of \cite{amendola}),
and also when  $w_1=w_2$. 
Now depending on whether $w_1>w_2$ or $w_2>w_1$, $\xi$ dynamically becomes very small or very
large in the early universe, being 
\be
\xi (\tau ) = \xi_{in} \; a^{\frac{3(w_1-w_2)}{(1+3 w_2)}} \, .
\ee
In particular
\be
 3 \,w_1+1 -\frac{4 f_1}{f_2} =
\begin{cases}\frac{3 \,\left(3\, w_1+1\right) \left(w_2-1\right)}{3\,
    w_2+1} <0  & {\rm for}\quad \xi\to \infty \\[.2cm]
 3 \,(w_1-1) <0& {\rm for}\quad \xi \to  0 
\end{cases} \, .
\ee
Thus, (\ref{condC}) cannot be satisfied at early times. When $\xi \to
0$ or $\xi \to \infty$, the mass term of $\E$ becomes time independent
and negative definite, leading to an exponential instability.
 We conclude therefore that also in the case ({\bf C}) the
instability cannot be avoided if $w_2 <1$.  For what concernes super
horizon modes the discussion is similar to the case ({\bf B}).

\section{Conclusions}
\label{con}
We studied in detail the dynamics of scalar perturbations in massive bigravity. Beside
its theoretical interest, massive gravity could be an interesting alternative
to dark energy.
 As a general ground, the ghost free massive gravity theories can be classified
  according to the global  symmetries of the potential $V$ in the unitary gauge~\cite{uslong}. 
 The ones characterized by Lorentz invariance on flat space have a number of issues
 %  Pauli-Fierz-like massive gravity theories 
 once an homogeneous FRW background is implemented . 
 %when applied to  cosmology.
 
   In  the bigravity formulation, with a \underline{single  matter sector},  things get better and FRW cosmological solutions indeed exist~\cite{russ,uscosm,hasscosm}. 
However, cosmological perturbations are different from the ones in GR. 
Already during radiation domination, sub horizon scalar perturbations tend to
grow exponentially~\cite{uspert}. 
The manifestation of such instabilities is rather peculiar. 
In the sector one, composed by ordinary matter and the metric $g$, their
perturbations are very close to the ones of GR.
The instability manifests as an exponential sub horizon growth of the field $\E$ and of the
second scalar mode $\Phi_2$, one of the Bardeen potentials of $\tilde
g$, which quickly invalidate the use of perturbation theory  at very early time.
This is very different from GR where 
perturbations become large (power law growth) only when the universe
is non relativistic. 

The emergence of an instability only in the perturbations of the second metric 
suggests its origin may resides in the  matter  content asymmetry of
the two sectors, since only the physical metric is coupled to 
matter. Indeed,  the only background solutions acceptables have
a ratio $\xi=\omega/a$  of the metrics' scale factors such that $\xi \ll 1$.

Adding a \underline{second matter sector} sourcing the second metric, opens up the
possibility (case ({\bf C})) to have a more symmetric background  with
$\xi  \sim 1$ and one may hope the exponential instability to be absent.  
Unfortunately, we have shown that this is not the case. 
Though, the pressure provided by the second matter stabilizes $\Phi_2$ and its dynamics becomes similar to GR,  the sub horizon
instability persists for $\E$ that represents a purely gravitational extra 
scalar field. 

We managed to analyze the perturbations  in whole range of  $\xi$
compatible with the early Universe evolution (matter and radiation).
The cases  ({\bf A}) and ({\bf B}) represent regions of very small $\xi$
where only one matter sector dominates, likewise the case with a single
matter, and both $\E$ and $\Phi_2$  grow exponentially inside the horizon.
When $\rho_1 \gg \rho_2$, the values of the tachyonic mass responsible
for that instability does not depend on $w_2$ and actually 
coincides with the one found in the case where $\rho_2=0$~\cite{uspert}.
In region ({\bf C}) both the matter sectors are important. 
 While, the Bardeen potentials
$\Phi_{1,\,2}$ are stable, the purely scalar gravitational
field $\E= E_1-E_2$ (see Appendix \ref{pert-app}) that involves both
metrics has  early time instabilities.
Finally, the region  ({\bf D}), characterized by very large values of
$\xi$, already at the level of background, spoils early time standard FRW cosmology.

Spanning the whole range of $\xi$ 
compatible with a standard early time cosmology, when $m^2 M_{pl}^2$ is
the order of the present cosmological constant, the bottom line is that 
massive bigravity has an intrinsic exponential instability. 

Looking at the behaviour of the matter contrast which  is the same of GR, one may speculate that some sort of Vainshtein~\cite{vain} cosmological mechanism could take place,
though here the trouble is with perturbations and not with the background.
Even if that happens, the deal is rather pricey: perturbation theory will
fail both at Solar System and cosmological scales.

\vskip 1cm
\no
{\Large \bf Acknowledgements}
\vskip .5cm
\no
M.C. thanks A. Emir G\"umr\"uk\c c\"uo\u glu for useful discussions
and the {\it Fondazione Angelo Della Riccia} for financial support.
L.P. thanks the Cosmology and Astroparticle Physics
Group of the {\it University of Geneva} for hospitality and support.
D.C.  thanks  { Negramaro} for their  {\it Senza fiato}    inspiring song. 

\begin{appendix}

\section{Perturbed Geometry}
\label{pert-app}
Let us now consider the perturbations of the FRW background (\ref{frw})
\be
g_{\mu \nu} = \,  \bar g_{1 \, \mu \nu}  +a^2 \,  {h_1}_{\mu
  \nu} \, , \qquad {\tilde g}_{\mu \nu} =  \bar g_{2 \, \mu \nu}  +
 \o^2 \,  h_{2 \, \mu \nu}  \, .
\ee
parametrized as follows 
\be
\begin{split}
& {h}_{1 \, 00} \equiv - 2 A_1  \, , \qquad  {h}_{2 \, 00}  \equiv - 2 c^2 \, A_2\\
& {h}_{1/2 \, 0 i}  \equiv {\cal C}_{1/2 \,i} - \de_i B_{1/2}  \, , \qquad
\de^i {\cal V}_{1/2 \, i}=  \de^i {\cal C}_{1/2 \, i} = \de^j {h^{TT}}_{1/2 \, ij} =
\delta^{ij} {h^{TT}}_{1/2 \, ij}  =0  \, ,\\
& h_{1/2 \, ij}  \equiv {h^{TT}}_{1/2 \, ij} + \de_i {\cal V}_{1/2 \, j} + \de_j {\cal
  V}_{1/2\, i} + 2 \de_i \de_j E_{1/2} + 2 \, \delta_{ij} \, F_{1/2}  \,.
\end{split}
\ee
Spatial indices are raised/lowered using the spatial flat metric.
In the scalar sector we can form 6 independent gauge invariant
scalar combinations  that we chose to be 
\be
\begin{split}
& \Psi_1= A_1 - {\cal H} \, \Xi_1   - \Xi_1^\prime \qquad \Psi_2= A_2 + c^{-2} \left(\frac{c'}{c} - {\cal H}_\o \right) \, \Xi_2   -
\frac{\Xi_2^\prime}{c^2}  \\
&\Phi_1 = F_1- {\cal H} \, \Xi_1 \, , \qquad \Phi_2 = F_2 - {\cal
  H}_\o \, \frac{\Xi_2}{c^2}  \, , \\
&{\cal E} = E_1 - E_2 \, ,  \qquad {\cal B}_1 = B_2 - c^2
B_1 +(1-c^2) \, E_1' \, ,
\end{split}
\label{sgib}
\ee
where $\Xi_{1/2} = B_{1/2} + E_{1/2}^\prime$.  In the matter sectors, we define the following gauge
invariant perturbed pressure and density
\be
\begin{split}
&\delta \rho_{1_{gi}} = \delta \rho_1 - \Xi_1 \, \rho_1' \, , \qquad \delta
p_{1_{gi}} = \delta p_1 - \Xi_1 \, p_1' \; ; \\
&\delta \rho_{2_{gi}} = \delta \rho_2 - \frac{\Xi_2}{c^2} \, \rho_2' \, , \qquad \delta
p_{2_{gi}} = \delta p_2 -  \frac{\Xi_2}{c^2} \, p_2'  \, .
\end{split}
\ee
The  scalar part $v$  of the perturbed 4-velocity $u^\mu$ is defined as 
\be
\begin{split}
&u_{1/2}^\mu = {\bar u}_{1/2}^\mu  + \delta u^\mu \, , \qquad u_1^\mu\; u_1^\nu\; g_{\mu
  \nu} = -1 \, , \quad u_2^\mu \;u_2^\nu\; \tilde g_{\mu
  \nu} = -1 \, , \quad  \delta u_{1/2}^0 = - a^{-1} \,  A_{1/2} \, ;  \\
& \delta u_{1/2 \, i} = a \;\left( \de_i v_{1/2} - \de_i B_{1/2}  \right) \, .
\end{split}
\ee
The corresponding gauge invariant quantity are defined as 
\be
u_{1/2 \,s} = v + E_{1/2}' \, .
\ee
The conservation of the EMT leads to a set of differential
relations. For the sector 1 we have
\bea
\delta \rho_{1_{ gi}}'=(1+w_1) \left[\rho_1 \,\left( k^2 \, u_{1 \, s}-3 \, \Phi_1'
  \right) - 3  \, {\cal H } \, \delta \rho_{1_{ gi}} \right] \, ,\\
u_{1 \, s}'= (3 w_1-1) \, u_{1 \, s} \, {\cal H} - \frac{w_1}{(1+w_1)} \, \frac{\delta
\rho_{1_{gi}}}{\rho_1} - \Psi_ 1\, .
\ena
For the sector 2
\be
\begin{split}
& \delta \rho_{2_{gi}}'=(1+w_2) \left[\rho_2 \,\left( k^2 \, u_{2 \, s}-3 \, \Phi_2'
  \right) - 3  \, {\cal H}_\o \, \delta \rho_{2_{gi}} \right] \,, \\
& u_{2 \, s}'= u_{2 \, s} \, \left[(3 w_2-1) \, {\cal H}_\o + \frac{c'}{c}\right]- c^2 \left[\frac{w_2}{(1+w_2)} \, \frac{\delta
\rho_{2_{gi}}}{\rho_2} + \Psi_ 2 \right] \, .
\end{split}
\ee
The perturbed Einstein equations for the first metric reads
\bea
&&2 \Delta  \Phi_1+6 {\cal H} \left(\Psi _1 \mathcal{H}-\Phi
  _1'\right) +a^2 m^2 \,  f_2 (3 {\cal F}_1-\Delta   {\cal
  E})  =-8 \pi  a^2 \,G \,\delta \rho_{1 _{gi}} \, ; \label{1tt}\\[.3cm]
&&\de_i \left[2 \Psi _1 \mathcal{H}-2 \Phi _1' +\frac{a^2 m^2  \,
    \mathcal{B}_1 \, f_2 }{(c+1)} +  8 \pi\,  G \,a^2\, (p_1+\rho_1 )\,
  u_{1 \,s}\right]
=0 \, ;  \label{1ts}\\[.3cm] 
&&  \left(\de_i \de_j - \delta_{ij} \Delta \right)
   \left( a^2 \,f_1\, m^2 \mathcal{E} - \Phi_1- \Psi_1\right)  + \delta_{ij}
   \left[m^2   \, a^2 \,(2\, f_1\, \mathcal{F}_1  +f_2\, \mathcal{A}_1 )+2
     \Psi_1 \,\left(\mathcal{H}^2+2 \,\mathcal{H}'\right) \right. \nb \label{1ss}\\[.2cm]
&&\left. -2 \,\Phi_1''-2\, \mathcal{H} \left(2\, \Phi _1'-\Psi _1'\right)\right]
= 8 \pi  G \, a^2 \, \delta_{ij} \,  
\delta p_{1_{gi}} \, , 
\ena
where
\be
f_1 =\xi \, \left[2 \,\xi  \,\left(3\, a_3 \, c \, \xi +a_2
 \,  (c+1)\right)+a_1\right] \, , \qquad f_2 =\xi \, \left(6 \,a_3 \,\xi ^2+4
\, a_2\, \xi +a_1\right) \, . 
\label{fdef}
\ee
For the metric $\tilde g$
\bea
&&
2\,c^2 \Delta  \Phi _2+6
   \mathcal{H}_{\omega } \left(\Psi _2   \mathcal{H}_{\omega }-\Phi _2'\right)  +\frac{m^2  a^2  f_2}{\kappa\,\xi^2}\, c^2 
\,   \left(\Delta  \mathcal{E}-3 \,\mathcal{F}_2\right)=-\frac{8 \pi \, G}{\kappa} \, a^2 \, c^2 \, \xi^2 \,\delta \rho_{2_{gi}} \, ; \label{2tt}\\[.3cm]
&& \de_i \left[ 2\, c \, 
   \left(\Psi _2 \mathcal{H}_{\omega }-\Phi _2'\right)-\frac{m^2  \, a^2 \, f_2}{\kappa\,\xi^2\,(1+c)} \, \mathcal{B}_2 
   +  \frac{8 \pi\,  G}{\kappa} \, c\, a^2 \, \xi^2 \, (p_2+\rho_2 )\,  u_{2 \, s} \right] =0 \, ;
 \label{2ts}\\[.3cm]
&&
-c \, \left(\de_i \de_j - \delta_{ij} \Delta \right) \left[ 
\frac{a^2 \, f_1 \, m^2}{\kappa\,\xi^2} \, \mathcal{E}+ \, c   \, \left(\Phi_2+\Psi
     _2\right)\right]+\delta_{ij}\left[ \frac{m^2 \,a^2}{\kappa\,\xi^2} (2\,  c\, f_1 \, \mathcal{F}_2+  f_2 \, \mathcal{A}_2 ) +\right.
\nb\\[.2cm]
&&
\left.
2\left( \mathcal{H}_{\omega }^2+2\,\mathcal{H}_{\omega }'-2\,\frac{c'}{c}\, \mathcal{H}_{\omega }\right) \Psi _2-
2 \Phi_2''+2\left(\frac{c'}{c}-2\,\mathcal{H}_{\omega }\right)\, \Phi
_2'+2\,\mathcal{H}_{\omega }\,\Psi _2' \right]=\frac{8 \pi \, G}{\kappa} \, a^2 \, c^2 \, \xi^2 \, \delta_{ij} \,  
\delta p_{2_{gi}} \,.\,\,\,\,\,\,\,\,\,\,\, \label{2ss}
\ena
The gauge invariant fields ${\cal F}_{1/2}$ can be expressed in terms
of $\Phi_{1/2}$, ${\cal B}_1$ and $\E$ 
by using 
\bea
&& {\cal H}_\o \, {\cal F}_2 - {\cal H} \, {\cal F}_1 = ( {\cal H} -
{\cal H}_\o ) (\Phi_1 - \Phi_2) \, ; \\[.2cm]
&& c^2( {\cal F}_2+  {\cal F}_1) =( {\cal B}_1 - {\cal E}) ( {\cal H} +{\cal H}_\o)
-2 c^2  (\Phi_1 - \Phi_2)  \, .
\ena
We often use the Fourier transform of
perturbations with respect to $x^i$, the corresponding
3-momentum is $k^i$ and $k^2 = k^i k_i$. To keep notation as
simple as possible we give up the symbol of the Fourier transform.
  
\section{Evolution of Perturbations}
\label{details}
In this Appendix we give the equations that govern the evolution of
the perturbations. We are interested in two two regimes: sub horizon modes with $k \;\tau 
\gg 1$ and super horizon ones for which $k \;\tau\ll 1$.

\subsection{Case ({\bf B})}
%\begin{itemize}
Remember that in this case we have $\xi\simeq \;\frac{a_1\,m^2}{8\,\pi\,G\;\rho_1\;\kappa}\ll1$  and that   the leading contribution for the
evolution equation of $\Phi_1$ is the same of GR.
For \underline{\it sub horizon modes}, $\Phi_2$ and $\E$ satisfy a system of coupled
equation

\be
\begin{split}
&
\Phi _2''+\frac{6 \left[9 \left(w_2+1\right) w_1^2+3 \left(5 w_2+7\right)
   w_1+4 w_2+14\right]}{\tau \left(3 w_1+4\right) \left(3
   w_1+1\right)} \, \Phi_2'
   +  k^2 \left[\left(3 w_1+1\right) \left(3 w_1+4\right) w_2-3 w_1-2
     \right] \, \Phi _2\\
&+ k^2 (3 w_1-1) \, \Phi_1+
\frac{18 w_1-6}{\tau \left(3 w_1+1\right)} \, \Phi_1' \,+\frac{12 \left[9 \left(3 w_2+1\right) w_1^2+15 w_1 \left(3
   w_2+1\right)+2 \left(6 w_2+5\right)\right]}{\tau^3 \left(3
   w_1+1\right){}^3 \left(3 w_1+4\right)} \, \E' +\\
   & k^2\,\frac{6 \left(3 w_1+1\right) \left(3 w_1+4\right) w_2-6
   \left(w_1+1\right)}{\tau^2\, \left(3 w_1+1\right)^2} \,\E 
 =0 \; ; \\[.3cm]
& {\cal E}''+\frac{6 \left(9 w_2 w_1+9 w_1+12 w_2+10\right)}{\tau \left(9
   w_1^2+15 w_1+4\right)}  \mathcal{E}' +k^2 \left[w_1+3 \,\left(3\, w_1+4\right) \,w_2+1\right]\,
\mathcal{E}\\ & -\frac{1}{6}\, k^2\, \tau^2 \left(3 w_1-1\right) \left(3 w_1+1\right)^2 \, 
\Phi_1 +\tau(1 - 9  w_1^2) \, \Phi_1' +\frac{1}{6}\, k^2\, \tau^2 \left(3 w_1+1\right)^2 \left(12 w_2+w_1 \left(9
   w_2+3\right)+2\right) \, \Phi_2 \\
&+ \frac{\tau \left(3 w_1+1\right) \left[12 w_2+w_1 \left(9
   w_2+3\right)+2\right]}{3\, w_1+4}\, \Phi_2' =0 \; ; 
\end{split}
\ee 

For \underline{\it super horizon modes}, 
the coupled equations for $\Phi_2$ and $\E$ are given by
\be
\begin{split}
&\Phi _2''  +\frac{6 \left(3
   w_1+4\right) \left(w_2+1\right) }{\tau \left(3
   w_1+1\right)} \, \Phi _2' +\frac{6 
   \left(15 w_1+17\right) \left[\left(3 w_1+4\right)
   w_2+1\right]}{\tau^2 \left(3 w_1+1\right){}^2}\, \Phi _2+\frac{24}{\tau^3 \left(3 w_1+1\right){}^3} \, \mathcal{E}'\\
&+\frac{48 \left[12 w_2+w_1 \left(9 w_2-3\right)+2\right]
   }{\tau^4 (3 w_1+1)^4} \, \mathcal{E}-\frac{18 
   \left(3 w_1+5\right) \left(\left(3 w_1+4\right)
   w_2+1\right)}{\tau^2 \left(3 w_1+1\right){}^2} \, \Phi _1-\frac{18 \left[(3 w_1+4)
   w_2+1\right]}{\tau \left(3 w_1+1\right)} \, \Phi _1'=0 .
\end{split}
\ee

\be
\begin{split}
&\mathcal{E}''+\frac{2 \left(3 w_1+7\right) }{\tau
    \left(3 w_1+1\right)}\mathcal{E}'
    +\frac{36 \left(3 w_1^2+\left(6 w_2+5\right) w_1+8
   w_2+2\right)}{\tau ^2 \left(3 w_1+1\right){}^3} \mathcal{E}
-\tau  \left(36 w_2+3 w_1 \left(9 w_2-2\right)+1\right) \Phi _1'\\
&-\frac{9  \left(3
   w_1+5\right) \left(4 w_2+w_1 \left(3 w_2-1\right)\right)}{3 w_1+1}\Phi _1+
   \tau  \left(3
   \left(3 w_1+4\right) w_2+1\right) \Phi _2'+\\&
   \frac{3
    \left(9 w_1^3+9 \left(5 w_2+2\right) w_1^2+\left(111 w_2+14\right)
   w_1+68 w_2+7\right)}{3 w_1+1}\Phi _2=0
 \end{split}
\ee

\subsection{Case ({\bf C})}
For this case $
\xi\simeq \left(\frac{\kappa\; \rho_{1}}{\rho_{2}}
\right)^{1/2} = \,\xi_{in} \;
a^{\frac{3\,(w_1-w_2)}{1+3\,w_2}}$.
At the leading order in the $\epsilon$ expansion, $\Phi_2$ satisfies
the following equation that is valid for any $k \;\tau$
\be
\label{eq2}
\Phi _2'' +\frac{6 \left(w_2+1\right) }{\tau
   \left(3 w_2+1\right)}  \Phi _2'+k^2 \frac{ \, w_2 \left(3 \, w_1+1\right){}^2}{\left(3
   w_2+1\right){}^2}\Phi _2=0 \, ;
\ee
For $\E$,  inside
the horizon, we get 
\be
\begin{split}
&
\mathcal{E}'' +\frac{2 \left[2 \, f_1  \left(3 w_2+1\right)+f_2
   \left(1-9 \, w_1 w_2\right)\right]}{\tau \, f_2  \left(3 \,
   w_1+1\right) \left(3 \, w_2+1\right)} \, \E'+\frac{k^2 \left[(3
  w_1+1)f_2 -4f_1 \right]}{3 f_2 
   (3w_2+1)} \, \E + \\
   & k^2\,\tau^2 \,\frac{\left(3 w_1+1\right){}^2 \left(f_2 \left(3
   w_1+1\right)-2 f_1\right)}{6 f_2 \left(3
   w_2+1\right)} \, \Phi_2+\tau \left(-\frac{2 f_1}{f_2}+3
 w_1+1\right) \, \Phi_2'\\
&-\frac{\tau \left(3 w_1+1\right) \left[f_2(3 
   w_1+1)-2 f_1\right]}{f_2 (3w_2+1)} \, \Phi_1' -k^2\,\tau^2 \,\frac{\left(3 w_1+1\right){}^2 \left(f_2 \left(3
   w_1+1\right)-2 f_1\right)}{6 f_2 \left(3
   w_2+1\right)} \,  \Phi_1=0
\end{split} \, .
\ee
The quantities $f_1$ and $f_2$ are defined in
(\ref{fdef}). Imposing that all mass terms are positive we get
precisely condition (\ref{condC}).
Notice that for $w_2=w_1$ the equation  for ${\cal E}$
reduces to 
\be
\mathcal{E}''-
\frac{6\left(w_1-1\right)}{\tau \, (3 w_1+1)}
\mathcal{E}'+\frac{k^2 \left(w_1-1\right)}{3 w_1+1} \mathcal{E}+{\cal
  F}(\Phi_1,\, \Phi_2)=0 \; , 
\ee
and the exponential instability is present.

\end{appendix}

\end{document}